\documentclass[fleqn,usenatbib]{mnras}

\usepackage{newtxtext,newtxmath}

\usepackage[T1]{fontenc}

\DeclareRobustCommand{\VAN}[3]{#2}
\let\VANthebibliography\thebibliography
\def\thebibliography{\DeclareRobustCommand{\VAN}[3]{##3}\VANthebibliography}

\usepackage{graphicx}	
\usepackage{amsmath}	
\usepackage{threeparttable}
\usepackage{ae,aecompl}
\usepackage{color,soul}
\usepackage{lscape,rotating}
\usepackage{hyperref}
\usepackage[svgnames]{xcolor}

\def\asec{\ifmmode ^{\prime\prime}\else$^{\prime\prime}$\fi}
\def\msun{M$_{\odot}$}
\def\rsun{R$_{\odot}$}

\def\degs{\ifmmode ^{\circ}\else$^{\circ}$\fi}
\def\amin{\ifmmode ^{\prime}\else$^{\prime}$\fi}
\def\asec{\ifmmode ^{\prime\prime}\else$^{\prime\prime}$\fi}

\def\psr{J1641}
\def\ergs{erg~s$^{-1}$}

\def\hip{HiPERCAM}
\def\fermi{\textit{Fermi}}
\def\eros{\textit{eROSITA}}

\newcommand{\flux}{erg~s$^{-1}$~cm$^{-2}$}

\usepackage{ulem}

\title[The black widow pulsar J1641+8049]{
The black widow pulsar J1641+8049 in the optical, radio and X-rays}
\author[Kirichenko et al.]{
A. Yu. Kirichenko$^{1,2}$\thanks{E-mail: aida@astro.unam.mx},
S. V. Zharikov$^1$,
A. V. Karpova$^2$,
E. Fonseca$^{3,4}$, 
D. A. Zyuzin$^2$, \newauthor
\ Yu. A. Shibanov$^2$, 
E. A. L\'opez$^{5,6}$,
M. R. Gilfanov$^{7,8}$, 
A. Cabrera-Lavers$^{9,10}$,
S. Geier$^{9,10}$, \newauthor
\ F.~A. Dong$^{11}$,
D.~C. Good$^{12}$,
J.~W. McKee$^{13,14}$,
B.~W. Meyers$^{15,11}$,
I.~H. Stairs$^{11}$, \newauthor
\ M.~A. McLaughlin$^{3,4}$,
J.~K. Swiggum$^{16}$
\\
$^1$Instituto de Astronom\'ia, Universidad Nacional Aut\'onoma de M\'exico, Apdo. Postal 877, Baja California, M\'exico, 22800 \\
$^2$Ioffe Institute, 26 Politekhnicheskaya, St. Petersburg, 194021,  Russia \\
$^3$Department of Physics and Astronomy, West Virginia University, PO Box 6315, Morgantown, WV 26506, USA \\
$^4$Center for Gravitational Waves and Cosmology, West Virginia University, Chestnut Ridge Research Building, Morgantown, WV 26505, USA\\
$^5$Instituto de Física, Universidad Nacional Autónoma de México, POB 20-364, Cd.Mx. 01000, M\'exico\\
$^6$Instituto de Investigación en Ciencias Físicas y Matemáticas, USAC, Ciudad Universitaria, 01012, Zona 12, Guatemala \\
$^7$Space Research Institute, Russian Academy of Sciences, Profsoyuznaya 84/32, 117997 Moscow,  Russia \\
$^8$Max-Planck-Institut f\"ur Astrophysik, Karl-Schwarzschild-Str. 1, D-85741 Garching, Germany \\
$^9$Instituto de Astrof\'isica de Canarias, V\'ia L\'actea s/n, E38200, La Laguna, Tenerife, Spain\\
$^{10}$GRANTECAN, Cuesta de San Jos\'e s/n, E-38712, Bre\~{n}a Baja, La Palma, Spain\\
$^{11}$Department of Physics and Astronomy, University of British Columbia, 6224 Agricultural Road, Vancouver, BC V6T 1Z1 Canada\\
$^{12}$Department of Physics and Astronomy, University of Montana, 32 Campus Drive, Missoula, MT\\
$^{13}$E.~A. Milne Centre for Astrophysics, University of Hull, Cottingham Road, Kingston-upon-Hull, HU6 7RX, UK\\
$^{14}$Centre of Excellence for Data Science, Artificial Intelligence and Modelling (DAIM), University of Hull, Cottingham Road, Kingston-upon-Hull, HU6 7RX, UK\\
$^{15}$International Centre for Radio Astronomy Research (ICRAR), Curtin University, Bentley, WA 6102, Australia\\
$^{16}$Department of Physics, Lafayette College, Easton, PA 18042, USA\\
}

\date{Accepted XXX. Received YYY; in original form ZZZ}

\pubyear{2023}

\begin{document}
\label{firstpage}
\pagerange{\pageref{firstpage}--\pageref{lastpage}}
\maketitle

\begin{abstract}
PSR J1641+8049 is a 2~ms black widow pulsar with the 2.2~h orbital period detected in the radio and $\gamma$-rays.  
We performed new phase-resolved multi-band photometry of PSR J1641+8049 using the OSIRIS instrument at the Gran Telescopio Canarias.
The obtained data were 
analysed 
together with the new radio-timing observations from the Canadian Hydrogen Intensity Mapping Experiment (CHIME), the X-ray data from the Spectrum-RG/\eros\ all-sky survey, and all available optical photometric observations. An updated timing solution based on CHIME data is presented, which accounts for secular and periodic modulations in pulse dispersion.
The system parameters obtained through the light curve analysis, including the distance to the source 4.6--4.8 kpc and the orbital inclination 56--59 deg, are found to be consistent with 
previous studies. 
However, the optical flux of the source at the maximum brightness phase faded by a factor of $\sim$2 as compared to previous 
observations. 
Nevertheless, the face of the J1641+8049 companion remains one of the most heated (8000--9500 K) by a pulsar among the known black widow pulsars. 
We also report a new estimation on the pulsar proper motion 
of $\approx$2 mas yr$^{-1}$, 
which yields a spin down luminosity of 
$\approx$4.87$\times 10^{34}$ ergs s$^{-1}$ and a corresponding heating efficiency of the companion by the pulsar of 
0.3--0.7.
The pulsar was not detected  in X-rays implying  
its  X-ray-luminosity was   
$\la$3 $\times$ 10$^{31}$ \ergs\ at the date of observations.

\end{abstract}

\begin{keywords}
stars: neutron -- binaries: general -- pulsars: individual: PSR J1641+8049
\end{keywords}

\section{Introduction}
\label{sec:introduction}

Among about 3400 pulsars discovered to date, 
more than 550 belong to the class of millisecond pulsars
(MSPs; \citet{atnf})\footnote{\url{https://www.atnf.csiro.au/people/pulsar/psrcat/}}. These objects have short spin periods ($P<30$ ms) and low spin-down rates ($\dot{P} \sim 10^{-20}$--$10^{-18}$ s~s$^{-1}$).
The most generally accepted scenario implies that MSPs are old 
neutron stars (NSs) which were spun-up (or `recycled') through 
angular momentum transfer by accretion from their main-sequence companions
during a low-mass/intermediate-mass X-ray binary stage
\citep{Bisnovatyi-Kogan1974,alpar1982}. 
The binary MSP population comprises several classes depending on the type 
of the companion. In the so-called `spider' systems with tight orbits 
($P_b \lesssim 1$ d), low-mass companions are heated and ablated by 
the pulsar wind of relativistic particles and radiation 
\citep[e.g.][]{manchester2017}. Evaporated material often causes 
eclipses of the pulsar radio emission. Black widows (BWs), which represent
a subclass of such binaries, have very low-mass ($M_c \lesssim 0.05$\msun)   
almost ablated degenerate companions. The origin and formation 
of such systems is not well understood, but it is actively discussed
\citep{chen2013,benvenuto2014,benvenuto2015,ablimit2019,ginzburg&quataert2021,guo2022}. 

About 70 BWs have been discovered so far thanks 
to radio and $\gamma$-ray observations. Half of them reside 
in the Galactic disk \citep{swihart2022}, while others are associated with globular
clusters\footnote{See \url{https://www3.mpifr-bonn.mpg.de/staff/pfreire/GCpsr.html}}. 
According to the recent census \citep{swihart2022}, only about 20 BWs have been detected in the optical.
However, optical studies allow one to determine fundamental parameters of BWs such as spectral type and temperature of a companion, irradiation efficiency, distance, as well as masses of its components 
when they cannot be derived from radio timing observations alone.

\begin{figure}

\begin{minipage}[h]{1.\linewidth}
\center{\includegraphics[width=1.0\linewidth, trim={0.5cm 1cm 1.1cm 1cm}, clip]{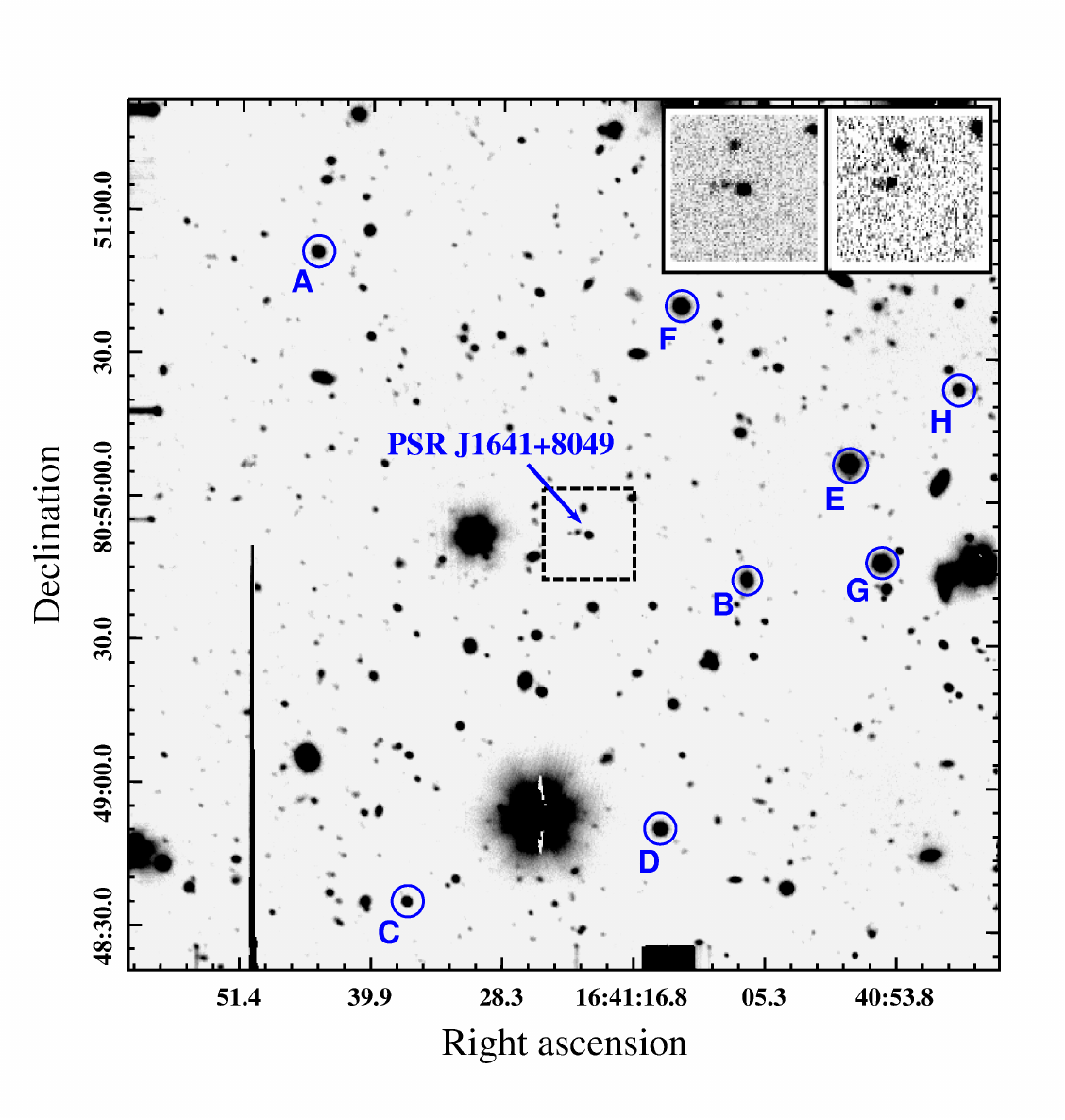}}
\end{minipage}

 \caption{The $3.05 \times 3.05$ arcmin$^2$ \psr\ FoV imaged with the GTC/OSIRIS in the $r'$ band. 
 The \psr\ vicinity is shown by the box in the centre of the image, and the optical companion is indicated by the arrow. The left and right inserts correspond to the enlarged \psr\ vicinity imaged near the companion maximum and minimum brightness phases.   
 The stars from the Pan-STARRS catalogue listed in Table~\ref{stars} and used for the photometric calibration, are marked by the capital letters. }  
	 \label{fig:field}
\end{figure}

\begin{table}
\caption{Log of the \psr\ observations with GTC/OSIRIS.}
\label{log}
\begin{tabular}{ccccccccc}\hline
Date & Filter& Exposure time,     & Airmass     & Seeing, \\ 
     &       & s                  &             & arcsec \\
\hline
12/07/2020 & $r'$ & 150$\times$43 & 1.62$-$1.70 & 0.8$-$1.0  \\
13/08/2020 & $g'$ & 200$\times$21 & 1.64$-$1.76 & 0.7$-$1.1 \\
13/08/2020 & $i'$ & 120$\times$21 & 1.64$-$1.76 & 0.7$-$0.9 \\
\hline
\end{tabular}
\end{table}

\begin{table}
\caption{Stars from the Pan-STARRS Catalogue detected in the \psr\ field and their magnitudes.}
\label{stars}
\begin{tabular}{ccccc}\hline
Star &  $g'$ & $r'$  & $i'$   \\ 
\hline
A & 21.619(068) & 20.575(025) & 19.598(017)  \\
B & - & 21.295(137) & 20.717(019)  \\
C & - & 21.316(044) & 20.636(032)  \\
D & 21.142(054) & 19.970(021) & 18.980(011)  \\
E & 19.617(014) & 18.559(009) & 18.054(008)  \\
F & 20.416(030) & 19.253(013) & 18.683(008)  \\
G & 19.297(012) & 18.744(006) & 18.505(008)  \\
H & 21.403(095) & 21.058(054) & 19.889(012)  \\

\hline
\end{tabular}
\end{table}

\begin{table}
\renewcommand{\arraystretch}{1.2}
\caption{The \psr\ parameters derived from observations with the CHIME/Pulsar backend.}
\label{tab:pars} 
\begin{center}
\begin{tabular}{lc}
\hline
\multicolumn{2}{c}{Logistics of CHIME/Pulsar radio observations}  \\
\hline
MJD Range of Observations, MJD                       & 59030--60063 \\
Frequency range, MHz                                & 400--800 \\
Number of channelized TOAs per epoch                & 32 (maximum) \\
Number of observing epochs                          & 581 \\      
Number of TOAs, total                               & 7793 \\
{\sc tempo} goodness-of-fit ($\chi^2$) statistic    & 9970 \\
{\sc tempo} reduced $\chi^2$ statistic              & 1.04 \\
{\sc tempo} root-mean-square residual, $\mu$s       & 1.88 \\
{\sc tempo} scaling factor for raw uncertainties    & 1.7 \\
\hline
\multicolumn{2}{c}{Best-fit timing parameters from {\sc tempo}}  \\
\hline
Right ascension (J2000), hh:mm:ss                   & 16:41:20.83311(3) \\
Declination (J2000), dd:mm:ss                       & 80:49:52.92335(8) \\
Proper motion in R. A., mas yr$^{-1}$               & $-$1.74(10) \\
Proper motion in Dec., mas yr$^{-1}$                & $-$1.03(11) \\
Timing parallax, mas                                & $-4(4)$ \\
Spin frequency $f$, Hz                              & 494.760636473171(4) \\
Time derivative in spin frequency, Hz$^2$ &  $-2.3928(2)\times10^{-19}$ \\
Epoch of reference for pulsar spin, MJD             & 59315.1255 \\
DM used for observations, pc cm$^{-3}$              & 31.091883 \\
Best-fit DMs, binned                                & see Section \ref{sec:chime} \\
Binary model used by {\sc tempo}                    & ELL1 \\
Projected semi-major axis, lt-s                     & 0.06407531(17) \\
Epoch of passage through asc.-node longitude, MJD   & 59315.04093360(5) \\
ELL1 eccentricity parameter \#1                     & 0.000019(4) \\
ELL1 eccentricity parameter \#2                     & $-0.000010(4)$ \\
Orbital frequency $n_b$, Hz                       & $1.2736404201(2)\times10^{-4}$ \\
First time-derivative in $n_b$, Hz$^2$              & $-9.9(3)\times10^{-20}$ \\
Second time-derivative in $n_b$, Hz$^3$              & $-1.89(16)\times10^{-27}$ \\
\hline
\multicolumn{2}{c}{Derived parameters from CHIME/Pulsar radio observations}  \\
\hline
Proper motion $\mu$, mas yr$^{-1}$                  & $2.02(10)$ \\
Observed spin-period derivative $\dot{P}$, s s$^{-1}$             & $9.7748(9)\times 10^{-21}$ \\
Intrinsic spin-period derivative $\dot{P}_i$, s s$^{-1}$             & $10.169(12)\times 10^{-21}$ \\
Mass function $f_M$, \msun\                         & $3.42 \times 10^{-5}$ \\
Characteristic age $\tau_c\equiv P/2\dot{P}$, Gyr                    & 3.28 \\
Observed spin-down luminosity $\dot{E}$, erg s$^{-1}$        & $4.67\times10^{34}$ \\
Intrinsic spin-down luminosity $\dot{E}_i$, erg s$^{-1}$        & $4.86\times10^{34}$ \\
Magnetic field at the equator $B_{\rm eq}$, G       & $1.42\times10^{8}$ \\
\hline
\end{tabular}
\end{center}
\begin{tablenotes}
\item Numbers in parentheses denote 1$\sigma$ uncertainties relating to the last 
significant digit quoted.
\item Spin-down luminosities are calculated assuming the canonical moment of inertia of 10$^{45}$ g cm$^2$. 
$\dot{E}_i$ is calculated based on the distance $D = 4.6(2)$ kpc derived from our modelling (see text).
\end{tablenotes}
\end{table}

The binary MSP PSR J1641+8049 (hereafter \psr) was discovered in the Green Bank 
North Celestial Cap (GBNCC) pulsar survey \citep{stovall2014,lynch2018}. 
This is an eclipsing radio pulsar which is also presented in the list of \fermi-detected pulsars \citep{fermipulsars}. 
Its flux in the 0.1--100 GeV range is $2.0(3)\times10^{-12}$~\flux\ \citep{4fgl-dr3}. \psr\ was not observed in X-rays.
It has an orbital period
of 2.18 h which is one of the shortest among the known BWs. The companion's minimum 
mass was estimated to be 0.04 \msun\ \citep{lynch2018}.  
The dispersion measure (DM) distances to the pulsar are $D_{\rm YMW16}$ = 3.0 kpc and $D_{\rm NE2001}$ = 1.7 kpc based on the YMW16 \citep{ymw2016} and
NE2001 \citep{ne2001} models for the distribution of free
electrons in the Galaxy, respectively.

\citet{lynch2018} found a faint optical counterpart ($r = 24.0(3)$) to \psr\ in the archival data, and performed photometric observations of the pulsar 
with the 4.3-m Lowell Discovery Telescope\footnote{Formerly known as the Discovery Channel Telescope} (LDT) in the $g$, $r$, $i$ and $z$ filters and with the McDonald Observatory 1 m telescope in the $r'$ and $i'$ filters. 
The counterpart revealed  strong brightness variations tied to the orbital period, confirming the optical identification of the pulsar companion.
Further optical studies of \psr\ were recently reported by \citet{matasanchez2023}
who performed its phase-resolved multiband photometry with the HiPERCAM 
instrument at the 10.4-m Gran Telescopio Canarias (GTC) 
in 2019.
They analysed the multiband light curves and obtained the fundamental parameters of the system, including the inclination, the Roche-lobe filling factor, the companion mass and the temperature gradient over  the companion surface, as well as the distance to the system.

In this paper, we report the results of our independent phase-resolved multi-band optical observations of \psr\ obtained with the OSIRIS instrument at the GTC in 2020. 
We analyse the OSIRIS, HiPERCAM, LDT, and the 1 m telescope data of \psr\ all together. 
In addition, we present the updated parameters of the system derived from ongoing observations with the Canadian Hydrogen Intensity Mapping Experiment (CHIME) telescope, and report an upper limit on the pulsar X-ray 
flux based on observations with \eros\ \citep{erosita2021} aboard the Spectrum-RG (SRG) orbital observatory \citep{Sunyaev2021}.
The paper is organised as follows: observations and data reduction are described in Sec.~\ref{sec:data}, 
the radio timing analysis is presented in Sec.~\ref{sec:chime},
while
the modelling of the light curves is described in Sec.~\ref{sec:lc-mod}. 
Discussion and conclusions are given in Sec.~\ref{sec:discussion}.

\section{Observations and data reduction}
\label{sec:data}

\subsection{Optical data}
\label{subsec:gtc}

The phase-resolved photometric observations\footnote{Proposal GTC11-20AMEX, PI A. Kirichenko} 
of the \psr\ field were carried out during two observing runs 
in the Sloan $g'$, $r'$ and $i'$ bands with the Optical System 
for Imaging and low Resolution Integrated Spectroscopy 
(OSIRIS)
instrument at the GTC. 
In order to reduce the CCD readout time and increase the efficiency of the phase-resolved observations, we used windowing. 
The target was exposed on CCD1, and the windowed FoV was $3.05 \times 3.05$ arcmin$^2$. To avoid effects from CCD defects, 5 arcsec dithering between the individual exposures was used 
in both observing runs. The observations roughly covered 
two orbital periods in total, i.e. one orbital period per each observing run. 
The first period was observed in the $r'$ band only, 
whereas the second one was covered one month later using the alternating $g'$ and $i'$ bands. 
The log of observations is given in Table \ref{log}. 
The $r'$-band image of the pulsar field is presented in Fig.~\ref{fig:field},  where the inserts demonstrate 
the variability of the pulsar companion. 

Using the Image Reduction and Analysis Facility ({\sc iraf}) package,
we performed a standard data reduction, 
including bias subtraction 
and flat-fielding. The cosmic rays were removed from all images 
with the L.A.Cosmic algorithm \citep{vandokkum}.
Astrometric referencing was performed using a single 180-s $r'$-band image and a set of stars from the
Gaia DR3 Catalogue 
\citep{2023A&A...674A...1G}.
Given the reduced image size in the windowing mode, only five field stars detected by Gaia appeared to be suitable for the astrometric purposes. Using these stars, we computed the astrometric solution with the formal rms uncertainties  $\Delta$RA~$\la$ $0.13$ arcsec 
and $\Delta$Dec~$\la$ $0.14$ arcsec.

Photometric calibration was performed using the Sloan photometric standards 
SA111-1925 
for the $r'$ band and PG1528+062B 
for the $g'$ and $i'$ bands \citep{smith} 
observed during the same nights as the target. Using their instrumental magnitudes 
and the site extinction coefficients $k_{g'}$ = 0.15(2), $k_{r'}$ = 0.07(1), and $k_{i'}$ = 0.04(1) \citep{extinc}, 
we calculated the zero points $Z_{g'}=28.53(2)$, $Z_{r'} = 28.94(1)$, and $Z_{i'}=28.42(1)$.
To verify these zero points, we checked the sky transparency variability by 
comparing the instrumental magnitudes of a field star during the observations of the target. 
The variation did not exceed errors of the flux measurements for the used star.
Nevertheless, since the standards and the target were observed at different sky positions, we also compared the magnitudes of several field stars with those from the Sloan Digital Sky Survey (SDSS) Release 14 Catalogue \citep{2018ApJS..235...42A} and the Pan-STARRS Catalogue \citep{flewelling2020}. 
To convert the Pan-STARRS magnitudes to the SDSS photometric system, we used the equation (6) from \citet{torny2012}. The stars used in this analysis are shown in Fig. \ref{fig:field} and listed in Table \ref{stars}.
Their catalogue $g'$-band magnitudes were found to be consistent within uncertainties with those calibrated using the photometric standard. However, in case of the $r'$ and $i'$ bands, we found a slight discrepancy between the respective magnitudes. In addition, we calculated the colour term corrections and found that they are negligible in the $r'$ and $i'$ bands, and only slightly affect the $g'$-band measurements. Accounting for all of the mentioned corrections, the resulting zero points are $Z_{g'} = 28.51(2)$,  
$Z_{r'} = 29.03(3)$, and $Z_{i'}=28.34(3)$. 
The point source $3\sigma$ upper limits in individual $g'r'i'$ exposures are $g'=26.4$, $r'=25.8$, and $i'=24.8$. We note that all magnitudes presented in this paper are in the AB system.

\subsection{Radio data}
\label{subsec:radio}

\psr\ is being observed by the CHIME telescope \cite[CHIME;][]{abb+22a}. For our present work, we processed and analysed high-cadence timing data recorded with the pulsar-timing backend built for CHIME \citep{abb+22b}. The CHIME/Pulsar backend generates folded profiles evaluated over 10-s integrations and 1,024 frequency channels that span the 400--800 MHz range. These data are coherently dedispersed in real time and prior to folding, using the value of DM listed in Table \ref{tab:pars}.

The CHIME/Pulsar data set on \psr\ currently spans $\sim$3 years of observations between early 2020 and 2023, and is being collected in support of ongoing GBNCC analyses \citep{msp+23}. All data were processed -- through statistical cleaning of radio-frequency interference and downsampling to 32 channels -- using the {\sc psrchive} \citep{vdo12} and {\sc clfd} \citep{mbc+19} analysis suites. Once initially cleaned, a subset of data was co-added to form a high-significance pulse profile that turned into a `standard template' after de-noising; this standard template was then used to compute times of arrival (TOAs) for the entire CHIME/Pulsar data set, yielding 32 TOAs per epoch that are each evaluated across 32 downsampled frequency channels. Any TOAs with S/N values less than 8.0, as determined with the {\sc pat} utility in {\sc psrchive}, were excluded from analysis based on sub-optimal detection statistics. For the remaining data set, a small, additional amount of TOAs ($\ll$1 per cent) were excised during the timing analysis due to corruption of the pulse profile from sub-threshold interference that was not detected during the data-preparation process.

\subsection{X-ray data}
\label{subsec:x-rays}

The \psr\ field was observed in the course of SRG/\eros\ all-sky survey in four visits spanning between April 9, 2020 and Oct. 13, 2021 with
the total exposure time of 2.8 ks. 
No source was statistically significantly detected at the pulsar position.
The derived upper limit on the 
unabsorbed flux is $\approx 1.3 \times 10^{-14}$
\flux\ 
 in the 0.5--10 keV range (90 per cent confidence), 
assuming a power law (PL) model with the photon index $\Gamma$~= 2.5
\citep[the average value for BWs,][]{swihart2022}
and the absorbing column density $N_{\rm H}=7 \times 10^{20}$  cm$^{-2}$.
The latter was derived using the reddening $E(B-V)=0.08$ mag obtained
for \psr\ from the extinction map of \citet{dustmap2019} and the empirical relation
from \citet{foight2016}. Note that the reddening is equal to that obtained
from the optical light curve modelling (see below).

\section{Updated Radio Timing of PSR J1641+8049}
\label{sec:chime}

We refined the timing model of \psr, based on the solution developed by \cite{lynch2018}, using the {\sc tempo} pulsar-timing package to obtain updated estimates of the spin, astrometric, and orbital parameters based on the CHIME/Pulsar data set alone. We also incorporated 355 additional degrees of freedom to fit for DM in contiguous time bins of 1.5-day extent that span the timespan of the data set. Due to the turbulent environments often observed in BW systems, we also explored the fitting of parameters that quantify variations in the orbital elements.

\begin{figure}
    \centering
    \includegraphics[scale=0.53]{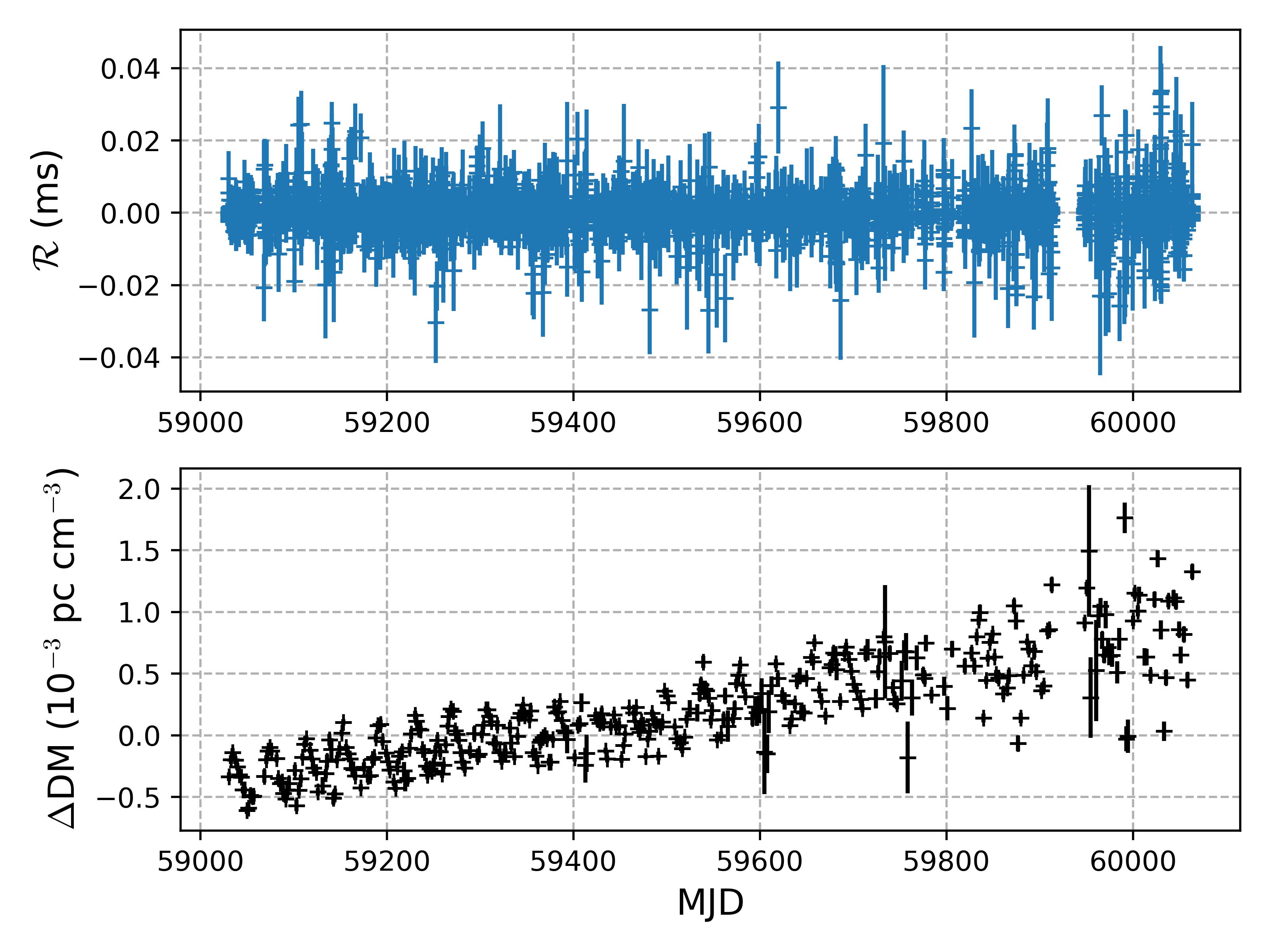}
    \caption{Best-fit timing residuals ($\mathcal{R}$, top) and DM values determined for the CHIME/Pulsar data set described in Section \ref{subsec:radio}. The DM values are plotted as changes relative to the value reported in Table \ref{tab:pars}.}
    \label{fig:J1641resids}
\end{figure}

A summary of the best-fitting CHIME/Pulsar timing residuals for \psr{} and DM timeseries is shown in Fig. \ref{fig:J1641resids}, and timing model parameters are reported in Table \ref{tab:pars}. One important result from our updated modelling is that the CHIME/Pulsar data yielded statistically different estimates of the proper motion than those obtained by \cite{lynch2018}, with the new magnitude of proper motion $\mu = 2.02(10)$ mas yr$^{-1}$ being lower by a factor of $\sim$20. We attribute this difference to the high-cadence nature of the CHIME/Pulsar TOAs, which allow for better estimates of short-term variations that are typically observed in other BW systems. This change in proper motion mainly impacts the derived estimate of `intrinsic' spin-down of the pulsar, i.e., the time rate of change in spin frequency corrected for biases induced by proper motion and acceleration in the Galactic potential \citep[e.g.,][]{nice1995}. The corrected spin-down we derive for \psr{} is discussed in Sec. \ref{sec:discussion} as it depends on results obtained from the optical analysis presented below. 

Our timing model of orbital motion uses the ELL1 formalism to describe low-eccentricity orbits \citep{lcw+01}. The best-fit ELL1 model indicates that deviations from purely-Keplerian motion are detectable in the CHIME/Pulsar data set. The inclusion of one time-derivative in orbital frequency $n_b = 2\pi/P_b$ as a degree of freedom improves the fit of the timing model by an amount $\Delta\chi^2 \approx 17,000$. We found that fitting for two time-derivatives in $n_b$, along with the five Keplerian elements, yielded an optimal fit to the timing data. The interpretation of these parameters in terms of macroscopic quantities (e.g., mass and/or geometry) is nontrivial due to the complex environment in BW systems that produce stochastic orbital variations \citep[e.g.,][]{shaifullah2016}. Future analysis of orbital evolution in the \psr{} system will be performed once several years of additional CHIME/Pulsar data are obtained.

The reliable detection of \psr{} indicates minimal eclipsing of radio signal at superior conjunction. However, the DM variations in Fig. \ref{fig:J1641resids} show secular and quasi-periodic trends over time. The estimate of $\mu$ for \psr{} nonetheless remains comparable whether we use a many-bin DM model or a polynomial expansion in DM, which differ in functional form and by hundreds of fit parameters. This circumstance indicates that proper motion is robustly measured with the CHIME/Pulsar data set despite only spanning $\sim$3 years in time and regardless of the choice in DM model.

In order to better assess these variations, we generated a separate timing solution based on data acquired in the MJD 59000--59400 range\footnote{While arbitrary, this range was chosen since the default version of {\sc tempo} is unable to accommodate the total number of per-epoch DM parameters needed to model the entire CHIME data set.} and setting maximum DM-bin extents to be 0.5 days, i.e., to ensure a single DM is estimated for each observing epoch. Despite the shortened data set, we fit for the same parameters reported in Table \ref{tab:pars} in order to separately measure variations in celestial position, DM and orbital motion.

\begin{figure}
    \centering
    \includegraphics[scale=0.53]{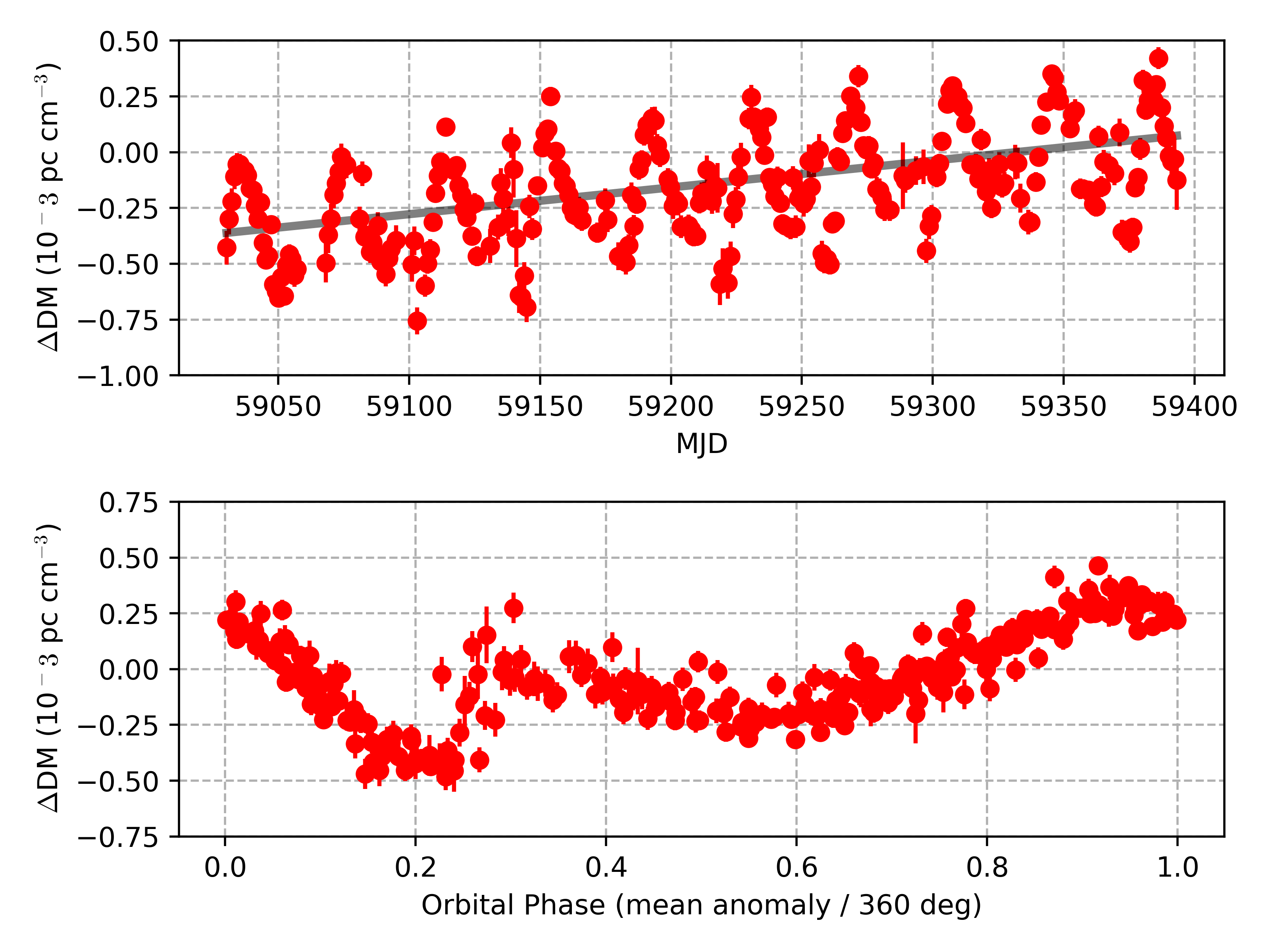}
    \caption{Per-epoch measurements of DM for \psr{} plotted as a function of time (top) and orbital phase (bottom), estimated by fitting to all CHIME/Pulsar data within the MJD 59000--59400 range. In the top panel, the grey line is a best-fit estimate of the secular variation over time that is presumed to be linear. In the bottom panel, the DM values have been corrected for the secular variation.}
    \label{fig:J1641DM}
\end{figure}

The results of this data-subset modelling is shown in Fig. \ref{fig:J1641DM}, which shows this residual subset as function of time and orbital phase. In the ELL1 binary model, the orbital phase $\Phi = n_b(t - T_{\rm asc})/(2\pi)$, where $T_{\rm asc}$ is the epoch of passage through the longitude of ascending node specified in Table \ref{tab:pars}; superior conjunction corresponds to $\Phi = 0.25$ in Figure \ref{fig:J1641DM}. The fitting of per-epoch DMs allows for better resolution of periodic variations, though requires at least 270 DM-bin fit parameters for the MJD 59000--59400 portion of the data alone. Nonetheless, the per-epoch DMs exhibit clear periodic variations that occur on a timescale equal to the orbital period for \psr; the $\sim$monthly variation in the DM timeseries is a manifestation of aliasing due to CHIME/Pulsar observations occuring once every sidereal day. While an excess in DM coincides with superior conjunction, the DM appears to modulate over the whole orbit and thus indicates a structured circumbinary medium. Further analysis of the DM variations over the entirety of the \psr{} data set, as well as other BW systems discovered and monitored by GBNCC, will be presented in future work.

\section{PSR J1641+8049 optical light curves and the system parameters}
\label{sec:lc-mod}

\begin{figure*}
\setlength{\unitlength}{1mm}
\resizebox{15.cm}{!}{
\begin{picture}(120,42)(0,0)
\put(57,0) {\includegraphics[width=7.5cm, bb = 0 220 1520 1100, clip=]{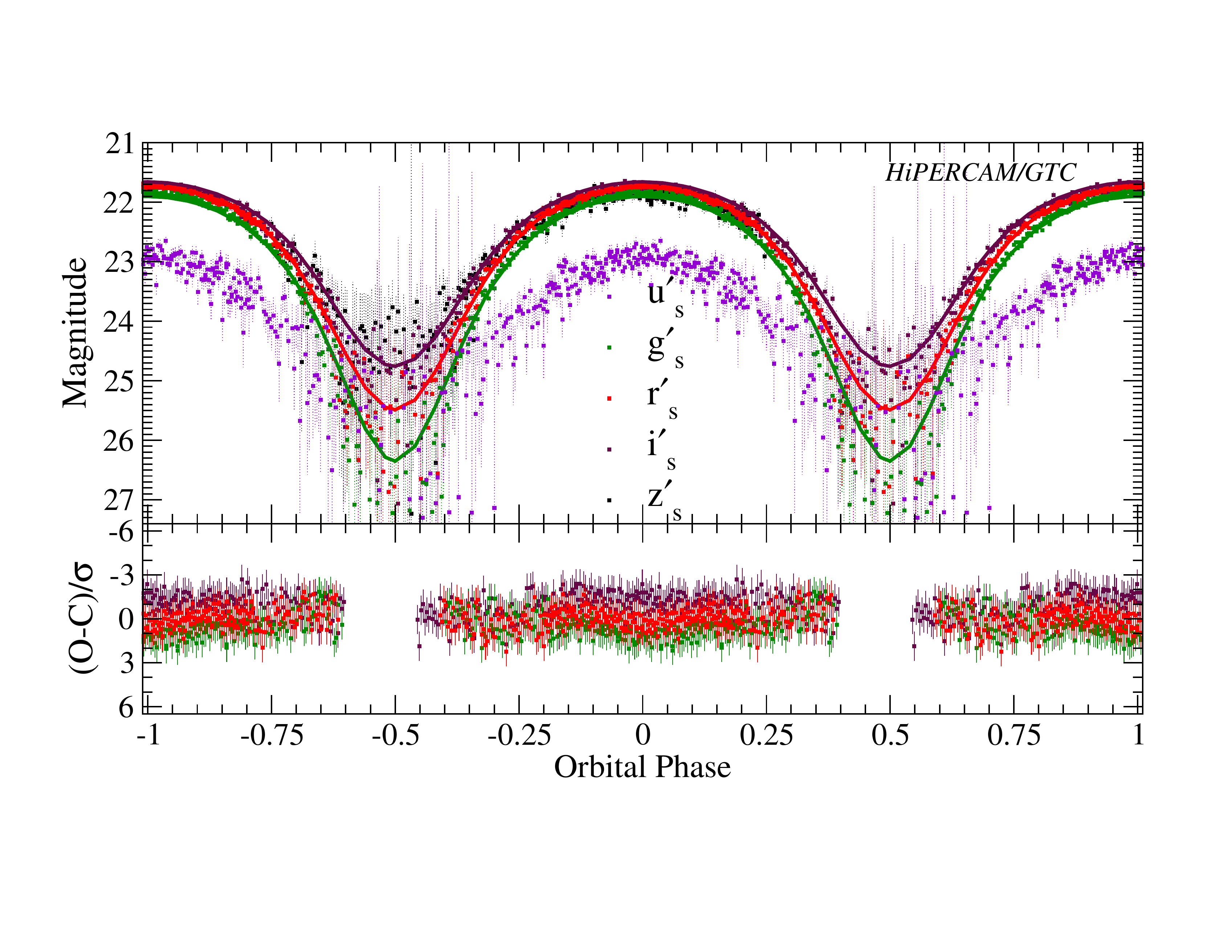}}
\put(-17,0) {\includegraphics[width=7.5cm, bb = 0 220 1520 1100, clip=]{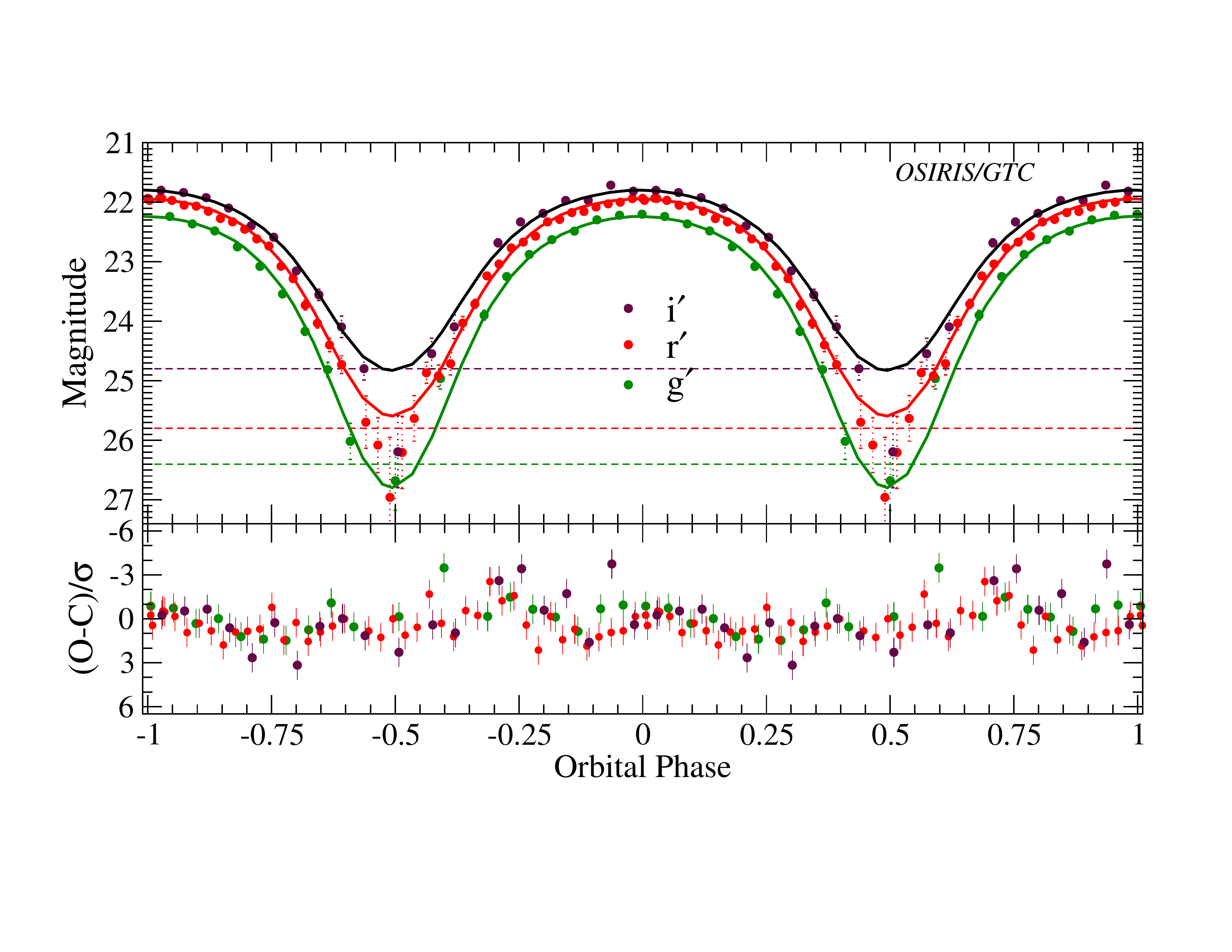}}
\end{picture}
}
 \caption{ \textit{Top-left}: Multi-colour light curves of J1641 obtained with the OSIRIS/GTC ($g'$, $r'$, $i'$) 
 folded with the orbital period. The phase zero corresponds to the orbital phase when the companion day-side is facing an observer. Two orbital cycles are shown for clarity. Solid lines represent the result of the best fit of the 
 data by  the model where the companion is heated by the pulsar. Horizontal  dashed lines show 3$\sigma$ detection 
 limits of the 
 observations. Photometric bands are marked by different colours. 
 \textit{Top-right}: The \hip/GTC multi-band data of J1641 and the best fit of the $g'_s$, $r'_s$, $i'_s$ light curves by the  model.
 \textit{Bottom panels}: Fit residuals calculated  as the difference between the observed ($O$) and the calculated ($C$) magnitudes for each data point   
 in terms of the magnitude error $\sigma$. }
	 \label{fig:psrLC}
\end{figure*}

\begin{figure}
\setlength{\unitlength}{1mm}
\resizebox{15.cm}{!}{
\begin{picture}(80,30)(0,0)
\put(0,0) {\includegraphics[width=7.5cm, bb = 0 30 1200 530, clip=]{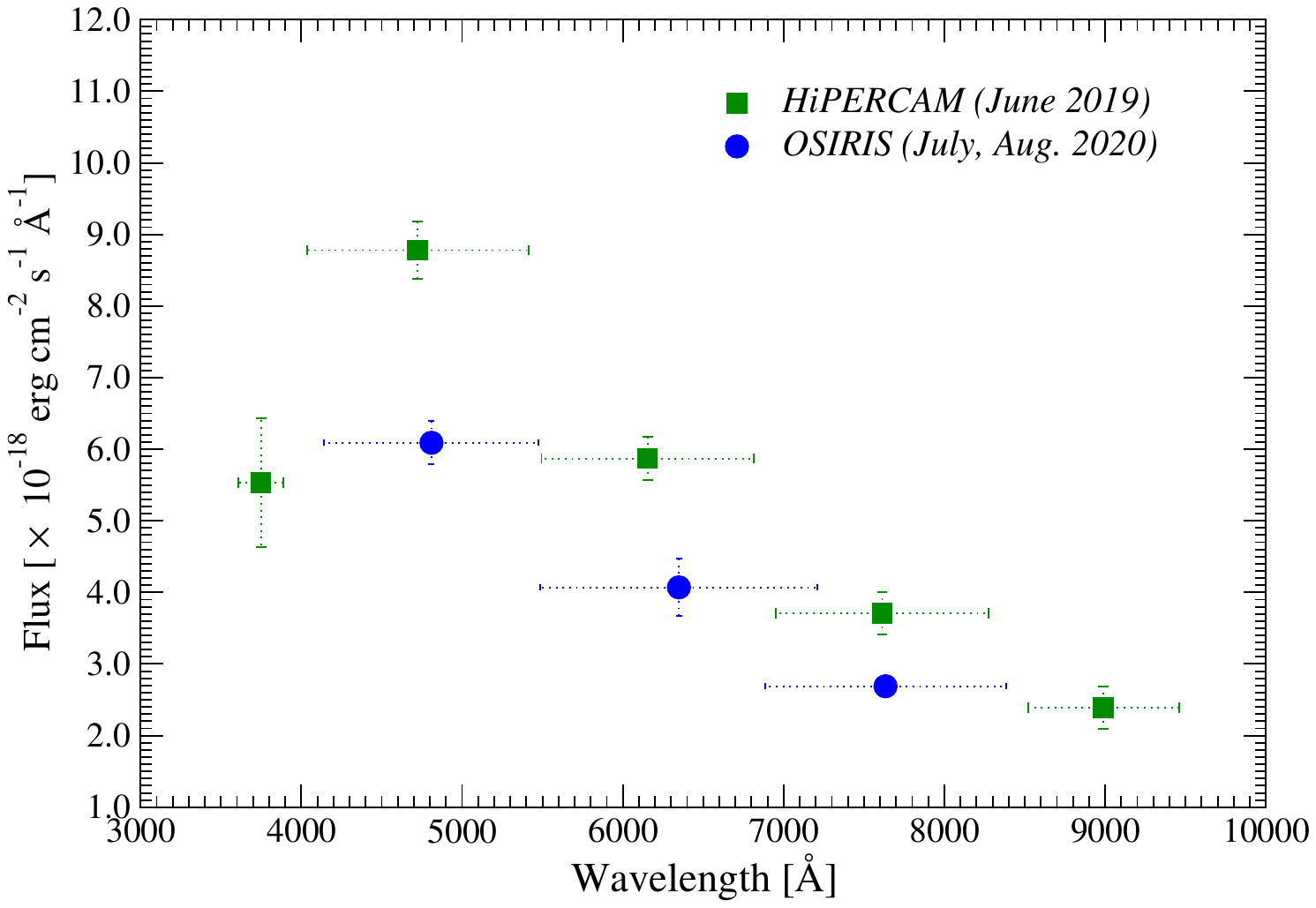}}
\end{picture}
}
 \caption{The broad-band optical spectra of the J1641 optical component at the photometric maximum for different epochs of observations.}

	 \label{fig:psrMax}
\end{figure}

\begin{figure}
\setlength{\unitlength}{1mm}
\resizebox{15.cm}{!}{
\begin{picture}(80,30)(0,0)
\put(0,0) {\includegraphics[width=7.5cm, bb = 0 30 1200 530, clip=]{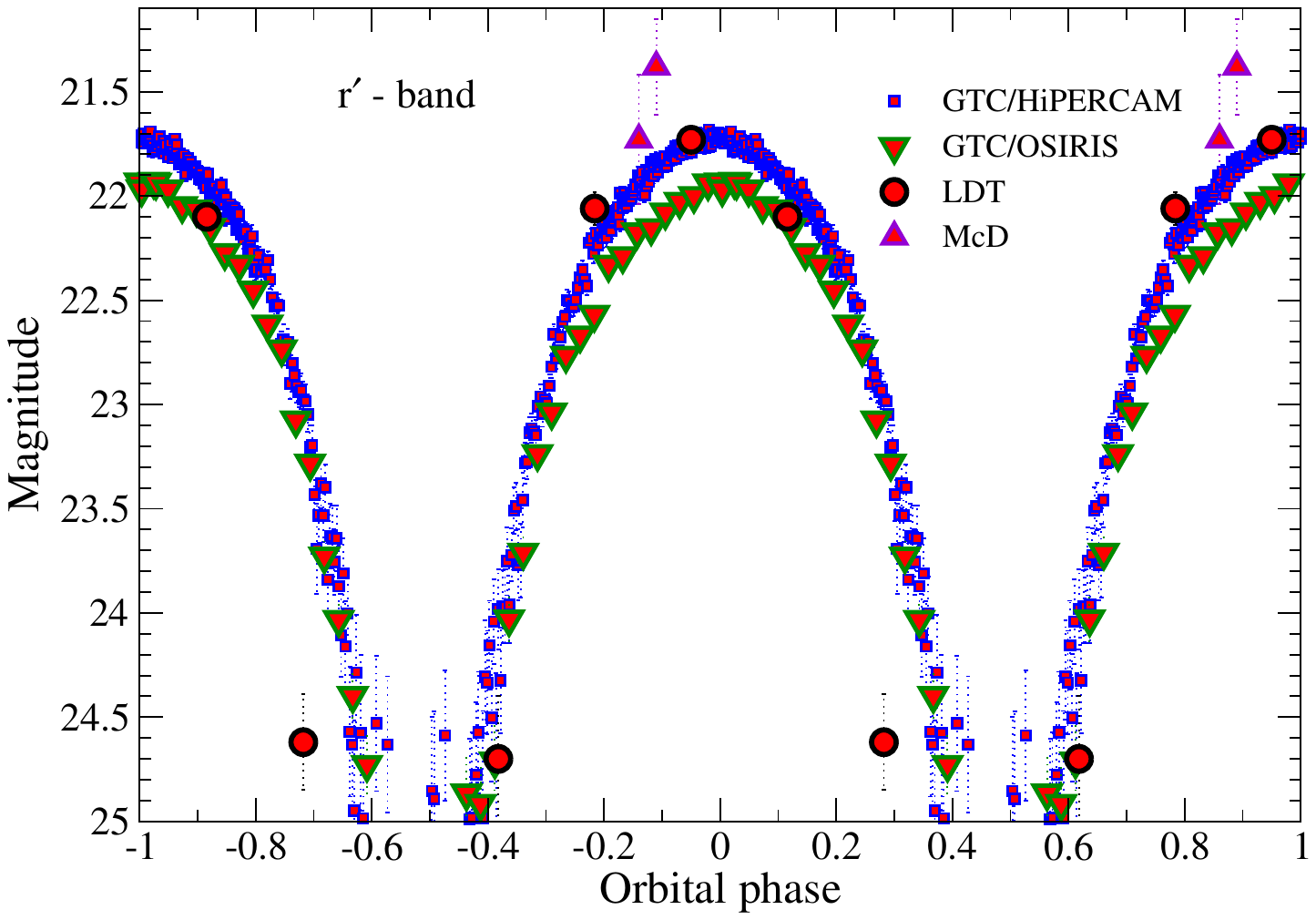}}
\end{picture}
}
 \caption{The r$^{\prime}$-band light curves folded with the orbital period obtained with the GTC, LDT, and McD. The orbital phase $\phi=0.0$ corresponds to the maximum of the GTC/HiPERCAM light curve and it is shifted to -0.76 from the radio convention orbital phase, where $\phi=0.0$ corresponds to the ascending node. LDT and McDonald 1-m telescope data were obtained in March 2017.}

	 \label{fig:rband}
\end{figure}

The resulting $g'$, $r'$, $i'$-band light curves folded with 
the orbital period 
are presented in 
Fig.~\ref{fig:psrLC}, left. 
In the right panel of Fig.~\ref{fig:psrLC} we  show the \hip\ data 
obtained in the $u'_s$-, $g'_s$-, $r'_s$-, $i'_s$-, and $z'_s$- bands about one year before our observations \citep{matasanchez2023}. The shapes of the light curves are found to be consistent, nevertheless the object appears to be slightly brighter and bluer at the maximum of the light curves in the \hip\ data.
To demonstrate this, 
in Fig.~\ref{fig:psrMax} we present the source broad-band optical spectra 
in the maximum of the light curves, corresponding to the OSIRIS (blue) and the \hip\ (green) data. 
The slopes\footnote{Without the $u'_s$ band for the \hip\ data.} of the spectra $F\sim\lambda^\alpha$ 
are $\alpha_{\rm OSIRIS} = -1.74(21)$ and $\alpha_{\rm HiPERCAM}=-2.01(18)$, respectively. 
As it can be seen, there is a brightness-decreasing tendency together with a relative reddening of the object spectrum on the time scale of these observations.
We also note that the flux measurements close to the photometric maximum obtained with the McDonald Observatory 1 m telescope 
were brighter 
than those reported for the GTC/HiPERCAM observations, 
whereas the LDT measurements were mostly close to them
(\citet[]{lynch2018} and D. Kaplan, private communication). For comparison, in Fig.\ref{fig:rband} we show the r$^\prime$-band light curves with all available data points.

\begin{figure}
\setlength{\unitlength}{1mm}
\resizebox{15.cm}{!}{
\begin{picture}(120,63)(0,0)
\put(0,0) {\includegraphics[width=6.0cm, bb = 50 30 1240 1350, clip=]{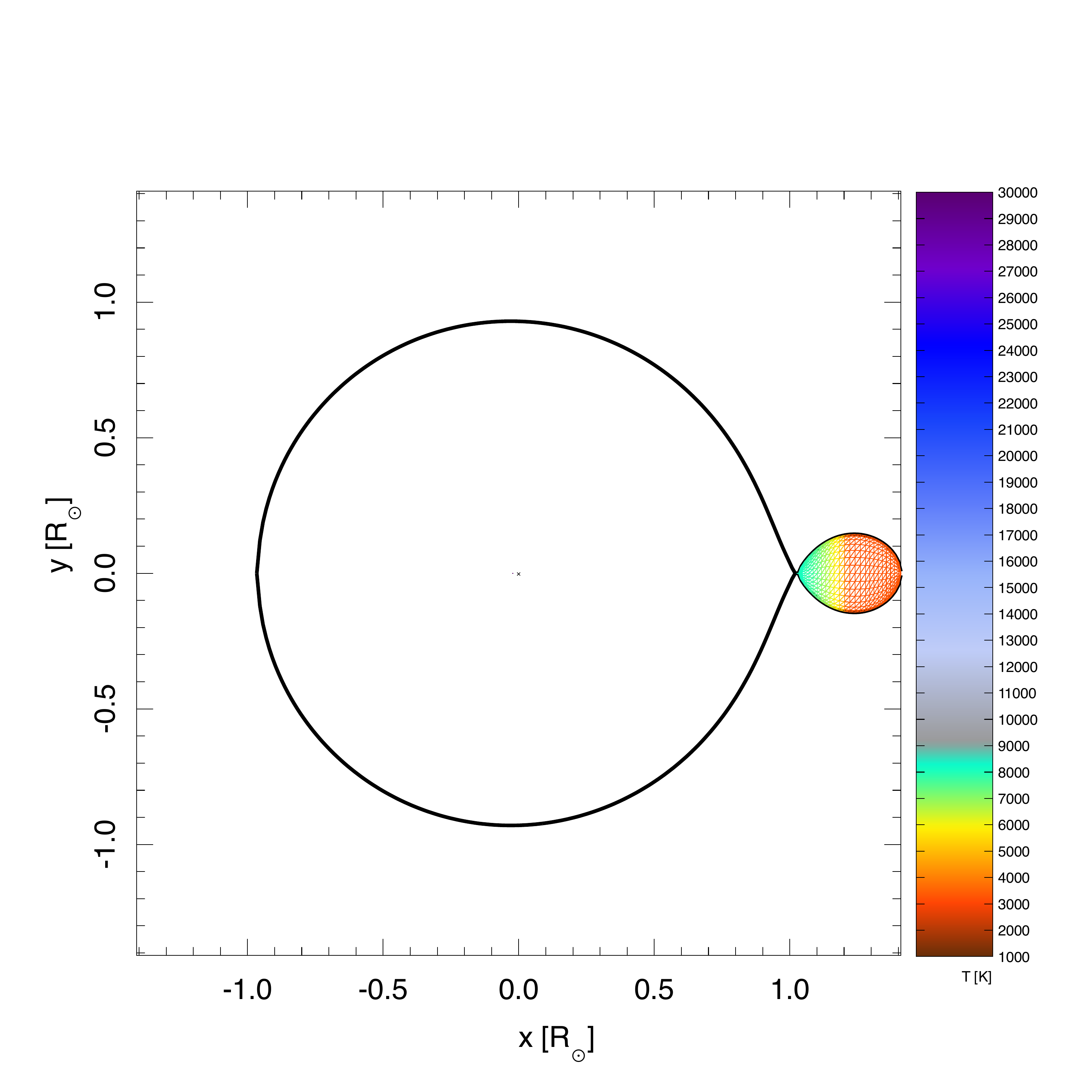}}
\put(60,0) {\includegraphics[width=1.01cm, bb = 1241 30 1440 1300, clip=]{Fig7real.pdf}}
\put(42,10) {\includegraphics[width=1.85cm, bb = 1095 654 1237.5 770, clip=]{Fig7real.pdf}}
\put(55,32){\vector(-1,-3){3}}

\end{picture}
}
 \caption{
 The Roche lobe of the system components and a magnification of the secondary.
 The colours mark the effective temperature distribution on the secondary surface. The image corresponds to the OSIRIS data modelling, when all parameters are free.
 }
	 \label{fig:psr2derr}
\end{figure}

\begin{table*}
\renewcommand{\arraystretch}{1.2}
\caption{The light-curve fitting results for \psr.}
\label{tab:fit} 
\begin{center}
\begin{tabular}{lcccc}
\hline

Mass function  $\times10^{-5}$, \msun\ (fixed)                                                    &  - &  -         &  3.42 & 3.42\\ \hline
Fitted parameters                                                   &  OSIRIS &  HiPERCAM  & OSIRIS &  HiPERCAM  \\

\hline
Pulsar mass $M_{\rm NS}$, \msun                                     & 2.0(6)   & 1.3(4)  &   {1.3(3)} &  1.3(1) \\
Mass ratio $q$ =  $M_{\rm c}/M_{\rm NS}$                            & 0.018(4) & 0.019(2)  &  0.035 & 0.036 \\             
Distance $D$, kpc                                                   & 4.60(20) & 4.70(20)  & 4.64(12) &  4.83(13) \\
Reddening $E(B-V)$, mag                                             & 0.072(23) & 0.056(30)  & 0.09(2) & 0.07(3)\\
`Night-side' temperature $T_{\rm n}$, K                             & 3300(100) & 3400$_{-300}^{+400}$  & 3380(120)  & 3500$_{-180}^{+300}$\\
Inclination $i$, deg                                                & 57(7)      & 56(10)     & 59(8)      & 58(10)\\
Roche lobe filling factor $f_x$                                     & 0.95$_{-0.10}^{+0.05}$ & 0.90$_{-0.09}^{+0.10}$  & 0.99$_{-0.08}^{+0.01}$ & 0.87$_{-0.07}^{+0.13}$  \\
Irradiation factor $K_{\rm irr}$,                                   &                         &     &                         &    \\
  $\times10^{20}$ erg~cm$^{-2}$~s$^{-1}$~sr$^{-1}$ &      $4.3(3)$                 &  5.8(5)    &        1.9(1)                &  3.6(2) \\
 $\chi^2$/d.o.f.         &           138/73              &  767/956   &           167/74              &  2525/957  \\
\hline
\multicolumn{5}{c}{Derived parameters}                                  \\
\hline
Companion mass $M_{\rm c}$, \msun                                   & 0.038 & 0.025   &      0.046                   &  0.047\\
Companion radius $R_{\rm c,x}$, \rsun                                 & 0.194 & 0.165  &     0.220                    & 0.198 \\
Companion radius $R_{\rm c,y}$, \rsun                                 & 0.143 &  0.121  &    0.167                     & 0.144 \\
Lowest `day-side' temperature $T_{\rm d}^{\rm min}$, K              & 3800 & 4250   &             3940            &4410 \\
Highest `day-side' temperature $T_{\rm d}^{\rm max}$, K              & 8200 & 9500  &             8000            & 8700 \\
 Irradiation efficiency $\eta$                                       & 0.45  &  0.7   &        0.3                 & 0.5 \\
\hline
\end{tabular}
\end{center}
\end{table*}

\begin{figure}
\setlength{\unitlength}{1mm}
\resizebox{15.cm}{!}{
\begin{picture}(80,35)(0,0)
\put(-2.5,0) {\includegraphics[width=7.75cm, bb = 0 30 1200 530, clip=]{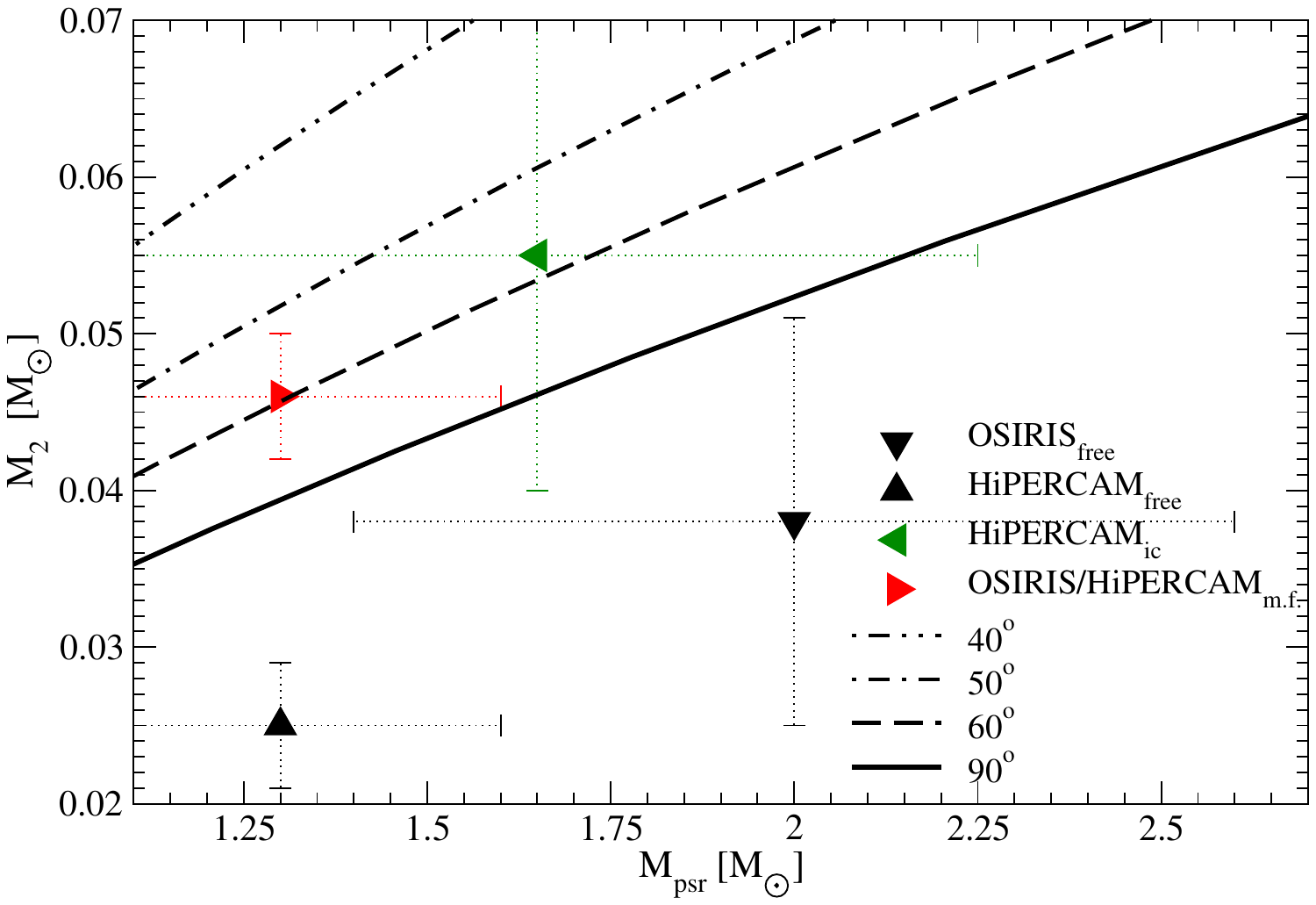}}
\end{picture}
}
 \caption{The masses of the pulsar and its companion from the fits of the optical light curves. 
 The lines show the relation between masses at different 
 orbit inclinations based on the mass function found from the pulsar radio timing.
 The black triangles  correspond to the fits when the masses and inclinations are free 
 parameters. The red triangle marks the fits which take into account the mass function. 
 The green  triangle shows the result from \citet{matasanchez2023}. }

	 \label{fig:mf}
\end{figure}


To estimate the system parameters, we fitted the light curves using the emission model of a binary system described in \citet{zharikov2013,zharikov2019}.  
The model consists of an NS as the primary which heats a low-mass companion as the secondary.
The spectrum  of each surface element of the companion is approximated by a blackbody with an effective temperature which is distributed non-uniformly over the star surface accounting for its heating by the pulsar.
The contribution of the pulsar into the observed optical flux is negligible for any expected distance and NS brightness values.
Following~ \citet{zharikov2019}, the effective irradiation factor of the secondary is related to the heating efficiency $\eta$ and the spin-down luminosity of the pulsar as 
\begin{equation}
K_{\rm irr} = \frac{\eta \dot{E_i}}{4\pi^2R_{\rm NS}^2},
\end{equation}
and it defines the effective radiate flux $F_{\rm in}$ transferred from the pulsar to the secondary:
\begin{equation}
    F_{\rm in} = \mathrm{cos}(\alpha_{\rm norm}) \Omega \Delta S K_{\rm irr},
\end{equation}
where $\alpha_{\rm norm}$ is the angle between the incoming flux and the normal to the surface,
$\Omega = \pi R^2_{\rm NS}/a^2$ is the solid angle from which the pulsar 
is visible from the surface element $\Delta S$ of the companion, $R_{\rm NS}=13$ km is the NS radius 
and $a$ is the orbit separation.
The corresponding `day-side' temperature of the companion star surface element is 
\begin{equation}
    T_{\rm d} = T_{\rm n}  \left[1 + \frac{F_{\rm in}}{\sigma \Delta S (T_{\rm n})^4} \right]^{1/4},
\end{equation}
where $\sigma$ is the Stefan-Boltzmann constant. The phase-resolved light curves were calculated by integrating the flux from all visible elements of the secondary in the corresponding band.
The gradient descent method was used to find the minimum of $\chi^2$  defined as
\begin{equation}
\chi^2 = \sum_j^{g',r',i'}\sum_k^{N_j} \left(\frac{O_{k}-C_{k}}{\sigma_{k}}\right)^2,
\end{equation}
where $N_j$ is the number of observations in a given filter, $O_{k}$, $C_{k}$, and $\sigma_{k}$ are the observed and the calculated magnitudes, and the error of the observed magnitude, respectively.
The free fitted parameters were the distance $D$, the reddening $E(B-V)$,  
the binary system inclination $i$, the Roche lobe filling factor $f_x$ defined as a ratio of distances from the centre of mass of the secondary to the star surface and to the Lagrange point $L_{1}$,  the `night-side' temperature $T_{\rm n}$ of the secondary, 
the effective irradiation factor $K_{\rm irr}$ [ergs s$^{-1}$ cm$^{-2}$ sr$^{-1}$], 
the pulsar mass $M_{\rm NS}$, and the component mass ratio.

The best-fitting model light curves for the OSIRIS data are shown in the left panel of  Fig.~\ref{fig:psrLC} by solid lines. The model parameters are given in the second column of Table~\ref{tab:fit}.  
The uncertainty of each fitted parameter was calculated following the method proposed by \citet{1976ApJ...208..177L}.
The geometry of the system and the distribution of 
the effective temperature at the companion surface for the OSIRIS data are shown in 
Fig~\ref{fig:psr2derr}.
For comparison, we also used the model for the HiPERCAM $g'_s$-, $r'_s$-, $i'_s$- band light curves. We limited our analysis of the HiPERCAM data to the three close optical bands because a simple blackbody spectrum approximation for radiation from a star surface element used in the model cannot   
describe the companion spectrum in a wide spectral range from the $u'_s$- up to the $z'_s$-band. On the other hand, this approach was applied to simplify the comparison of the fit results achieved using the same model setup in both cases.
Since the source was slightly brighter and bluer in the \hip\ data compared to the OSIRIS data, we first fixed all model parameters excluding the secondary heating at the values given in Table~\ref{tab:fit} for the OSIRIS data, and then fitted the \hip\ data. 
The fit provided a hotter day-side part of the secondary with a maximum temperature of about 9200 K.
After that, we thawed all parameters as in the case of the OSIRIS data. 
The best-fitting parameters of the last fit are  given in the third column of Table~\ref{tab:fit}.  
In general, they are close to or inside the 1$\sigma$ error range of the OSIRIS data fit results, except for the effective irradiation factor $K_{\rm irr}$, which, in turn, gives a higher day-side temperature of the secondary. Another difference is related to the mass of the pulsar. 
The fitting of the two data sets results in a significantly lower mass in the \hip\ data compared to that in the OSIRIS data. 
Nevertheless, their 1$\sigma$ uncertainties overlap. 
We note that the light curves shapes and fluxes  are mainly defined by the size of the companion Roche lobe,  its filling factor, temperature distribution and system inclination. 
The Roche lobe size is changed  by 15 per cent at the variation of the pulsar mass within its reasonable limits. To get the observed light curves, these changes can be compensated by variations of the temperature and filling factor. Thus, these parameters are correlated and additional information is needed to better constrain them. 

For instance, we can utilise the mass function from the new radio timing measurements 
(see Table~\ref{tab:pars}).
It decreases the numbers of free parameters, because it links the pulsar mass, the inclination and the mass of the companion.
 Taking into account the mass function   we repeated the fitting of the data from both instruments.  The results are presented in the last two columns of Table~\ref{tab:fit}. They are very close to 
 those obtained by \citet{matasanchez2023} who used the {\sc icarus} code \citep{2012ApJ...748..115B} to model the \hip\ data (see table~3 in the respective paper). However, we note that the formal
 $\chi^2$ values  
 are higher as compared to 
 the cases when the masses of components and inclination are free parameters. 
 All results are summarised in Fig.~\ref{fig:mf} and discussed below.

\section{Discussion and conclusions}
\label{sec:discussion}

As it was mentioned before, the \psr\ orbital period of 2.18~h is one of the shortest among the known BWs \citep{swihart2022}, and its highly modulated optical light curves are typical for such tight binary systems with MSPs. The shapes of the light curves do not demonstrate significant changes on a one-year time scale between the HiPERCAM and OSIRIS observations. However, the fluxes at the photometric maximum exhibit a significant decrease between the 
observations by up to a factor of two.       
This can indicate variability of the pulsar wind on a short (days) and long (years) time scales. 
We note that %
variable heating (increases and decreases of the companion irradiation by the pulsar) is observed for some other MSP 
binary systems, such as, 
e.g., the BW candidate 4FGL~J0935.3+0901 \citep{halpern2022} or the redback PSRs J1048+2339 \citep{yap2019} and J2129-0429 \citep{bellm}.
Follow-up observations are needed to confirm and study this effect for \psr.

The previously reported proper motion $\mu$ = $39(3)$ mas yr$^{-1}$ for \psr\
led to a negative value of the pulsar intrinsic spin-down luminosity, implying a spin-up scenario
\citep{lynch2018,matasanchez2023}, which could indicate the presence of accretion. 
However, the latter would complicate detection of radio pulsations from the pulsar and affect 
the optical light curves of the system.  
Neither the HiPERCAM nor the OSIRIS observations support this scenario. 
In addition, using the distance obtained through the HiPERCAM light-curve modelling, \citet{matasanchez2023} concluded that the maximum proper motion for this system, which would allow to avoid the spin-up scenario, is $\mu \leq 19$ mas yr$^{-1}$. 
Indeed, the updated proper motion $\mu$ = $2.02(10)$ mas yr$^{-1}$ derived 
from the new radio observations (see Sec.~\ref{subsec:radio}) is significantly lower than the one provided by \citet{lynch2018}. 
It yields a new estimation on the intrinsic
period derivative, $\dot{P_i}$ and the intrinsic spin-down luminosity 
$\dot{E_i}$. 
Considering the distance D = 4.6(2) kpc obtained from our optical light curve modelling, we estimated 
the Shklovskii correction \citep{shklovskii}, 
$\dot{P}_{\rm Shk}\approx 0.08\times10^{-21}$ s~s$^{-1}$.
 Taking into account the corrections due to the differential Galactic rotation and the pulsar acceleration, the corresponding intrinsic period derivative is $\dot{P_i}=10.2\times10^{-21}$ s~s$^{-1}$,  and the spin-down luminosity is $\dot{E_i}=4.87\times10^{34}$ \ergs, rejecting the spin-up and accretion scenario.
We note that in the case of this particular BW system, the Shklovskii correction is subdominant leading to the fact that the intrinsic spin-down luminosity is $\approx$ 4 per cent higher than the observed one.
In addition, taking into account the newly determined proper motion, we can now estimate the J1641 transverse velocity. Considering the distance 4.6 kpc derived from the light-curve modelling and following \citet{igoshev} to estimate the contribution of the Galactic rotation and solar peculiar velocity, we obtain $V_T \approx 36$ km~s$^{-1}$ in the local standard of rest of the pulsar. This velocity is typical for pulsar binary systems \citep{hobbs2005}.

According to \citet{matasanchez2023}, \psr\ has one of the heaviest companions ($M_{\rm c}$=0.055$^{+0.016}_{-0.014}$~\msun) among the known BWs. More massive secondaries
were found for, e.g., PSR J1555$-$2908 ($0.060^{+0.005}_{-0.003}$~\msun; \citealt{kennedy2022}) and PSR J1810+1744 (0.065(1)~\msun; \citealt{romani2021}). However, our fit for all free parameters resulted in a lower companion mass of 0.035(13)~\msun\ and 0.025(4)~\msun\ for the OSIRIS and \hip\ data, respectively. 
These values are close to the average mass $\bar{m}_c=0.026(15)$ of pulsar companions in BW systems  \citep[see table 5 therein]{swihart2022}. 
However, the value for the \hip\ fit is inconsistent with the mass function derived from the radio timing (see Fig.~\ref{fig:mf}), while the value for the OSIRIS fit agrees with it.
Taking the mass function into account 
we got the mass of the companion of 0.046(10)~\msun\ for both data sets. 
This is in agreement with the values reported by \citet{matasanchez2023}.
Thus, the companion indeed can be rather heavy in comparison with other BW systems.

\begin{figure}
\begin{minipage}[h]{1.\linewidth}
\center{\includegraphics[width=1.0\linewidth, bb = 00 00 410 325, clip=]{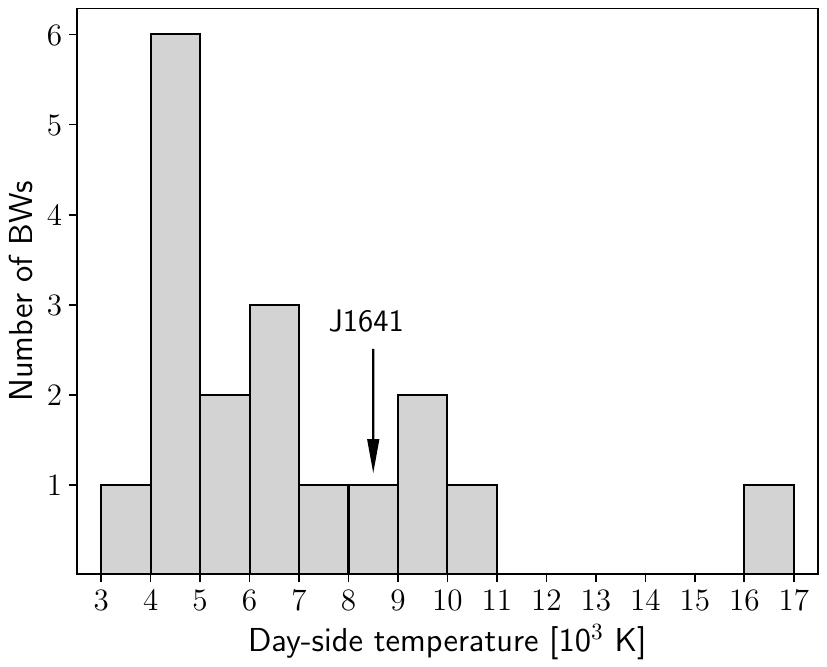}}
\end{minipage}
 \caption{Distribution of the day-side temperature $T_{\rm d}$ of BW companions. 
 18 sources are included.
 For PSRs J0023+0923, J0251+2606, J0636+5129, J0952$-$0607, J1124$-$3653,
 J1301+0833, J1544+4937, J1555$-$2908, J1653$-$0158, J1810+1744, J1959+2048,
 J2051$-$0827, J2052+1219, J2241$-$5236 and J2256$-$1024, the day-side temperatures
 are calculated using the base (night-side)  and the irradiation temperatures 
 ($T_{\rm n}$ and $T_{\rm irr}$) from \citet{matasanchez2023}  (see their table A1 and references therein).
 For PSR J1311$-$3430, \citet{kandel&romani2022} obtained very low $T_{\rm n}$,
 negligible in comparison with $T_{\rm irr}$, which was derived from 
 the irradiation luminosity and parameters of the orbit (i.e. $T_{\rm d}\approx T_{\rm irr}$).
 The position of \psr\ according to our fit of the OSIRIS data is marked by the arrow.
 We also include the BW candidate ZTF J1406+1222 \citep{burge2022}. 
 The best-fitting value of the base temperature of PSR J0610$-$2100 is 
 below the lowest temperature covered by the spectral models \citep{vanderwateren2022};
 thus, we excluded it from the sample.}  
	 \label{fig:hist}
\end{figure}

Another distinct feature of \psr\ is the high day-side temperature of the companion, 8200--9500~K,
making it one of the five most heated sources among the known BWs (see Fig.~\ref{fig:hist}).
The other four sources are PSR J1311$-$3430 with $T_{\rm d}\gtrsim12000$ K \citep{romani2012,romani2015}, the BW candidate ZTF J1406+1222 with $T_{\rm d}\approx10500$~K \citep{burge2022}, PSR J1810+1744 with $T_{\rm d}\approx9400$ K \citep{romani2021}
and PSR J1555$-$2908 with $T_{\rm d}\approx9400$ K \citep{kennedy2022}.
The \psr\ day-side temperature is $\gtrsim2$ times larger than the respective temperatures of, e.g., PSR J0251+2606 ($\approx$3400 K) or PSR J0636+5129 ($\approx$4600 K) \citep{matasanchez2023} near the low end of the source temperature distribution. It seems natural that the number of companions with a lower day-side temperature is larger than that with a hotter one, as the latter have to be  evaporated faster. However, it remains unclear how the companion heating is related to the ‘spin-down flux’ defined as $\dot{E}P_{b}^{-4/3}$ \citep[see figure 5 and table 5 therein] {zharikov2019}. It is possible that the absence of a clear dependence on the  spin-down luminosity and/or system separation indicates the importance of the pulsar wind instability and/or asymmetry in the companion heating.

Using the intrinsic spin-down luminosity $\dot{E_i}$ and the irradiation factor $K_{\rm irr}$, we estimated the irradiation efficiency range $\eta\approx$ 0.3--0.7. These values seem to 
overshoot those typically observed for BWs \citep[e.g.,][]{draghis2019}. 
An irradiation luminosity even larger than $\dot{E}$ ($\eta \ga 1 $), calculated assuming a canonical momentum of inertia value of 10$^{45}$ g~cm$^2$, was derived, e.g., for BW PSR J1810+1744 
by \citet{romani2021}. 
It is also an additional argument that
the true one can be lower due to the beaming factor \citep[see][]{draghis2019,romani2021}.

The derived companion mass of \psr\ (Table~\ref{tab:fit}) is close to  typical masses of brown dwarfs
(0.01--0.07~\msun) implying the possible origin of the companion.    
However, a field brown  dwarf with an age of $\ga 1$ Gyr, similar to the \psr\ characteristic 
age (Table~\ref{tab:pars}), would be twice more compact ($\la $0.1~R$_\odot$) 
and colder ($\la 1500$~K, \citealt{2021ApJ...920...85M}) as compared to the derived night-side 
temperature of \psr. This disfavours the  brown dwarf nature of the companion.    
Nevertheless, such a discrepancy  appears to be not a unique property of \psr.   
As was noted before, most BW companions, especially the strongly heated,
are bloated up by a factor of two  in comparison with the Galactic field brown dwarfs 
\citep[see table~6 therein]{kandel&romani2022}.
The strong irradiation by the pulsar  can affect the global structure of the presumed 
brown dwarf leading to the increase of its night side temperature as well. 

Finally, J1641 is not detected in X-rays in the 2.8 ks data accumulated during the SRG/\eros\ all-sky survey.
The upper limit (90 per cent) on its X-ray luminosity is about 3.3 $\times$ 10$^{31}$ \ergs\ 
for a distance of 4.6 kpc. This and the $\gamma$-ray 
luminosity $5\times10^{33}$~\ergs\ \citep{swihart2022} are typical for BWs.   

In addition to follow-up optical broad-band observations to confirm the optical variability of the companion, 
optical spectroscopy would be useful to establish its spectral type and to measure its radial 
velocity curve, to better constrain the parameters of the system.

\section*{Acknowledgements}

We thank the referee for their useful comments and suggestions which allowed us to improve the manuscript. We also thank Daniel Mata Sanchez and David Kaplan for providing the optical data and for useful discussions, and Andrei Igoshev for his helpful comments. 
The work is based on observations made with the Gran Telescopio Canarias (GTC), 
installed at the Spanish Observatorio del Roque de los Muchachos of 
the Instituto de Astrof\'isica de Canarias, on the island of La Palma. 
This work used data obtained with eROSITA telescope onboard SRG observatory. The SRG observatory was built by Roskosmos in the interests of the Russian Academy of Sciences represented by its Space Research Institute (IKI) in the framework of the Russian Federal Space Program, with the participation of the Deutsches Zentrum für Luft- und Raumfahrt (DLR). The SRG/eROSITA X-ray telescope was built by a consortium of German Institutes led by MPE, and supported by DLR.  The SRG spacecraft was designed, built, launched and is operated by the Lavochkin Association and its subcontractors. The science data are downlinked via the Deep Space Network Antennae in Bear Lakes, Ussurijsk, and Baykonur, funded by Roskosmos. The eROSITA data used in this work were processed using the eSASS software system developed by the German eROSITA consortium and proprietary data reduction and analysis software developed by the Russian eROSITA Consortium.
We acknowledge that CHIME is located on the traditional, ancestral, and unceded territory of the Syilx/Okanagan people. We are grateful to the staff of the Dominion Radio Astrophysical Observatory, which is operated by the National Research Council of Canada.  CHIME is funded by a grant from the Canada Foundation for Innovation (CFI) 2012 Leading Edge Fund (Project 31170) and by contributions from the provinces of British Columbia, Qu\'ebec and Ontario. The CHIME/FRB project is funded by a grant from the CFI 2015 Innovation Fund (Project 33213) and by contributions from the provinces of British Columbia and Qu\'ebec, and by the Dunlap Institute for Astronomy and Astrophysics at the University of Toronto. Additional support is provided by the Canadian Institute for Advanced Research (CIFAR), McGill University and the McGill Space Institute thanks to the Trottier Family Foundation, and the University of British Columbia. The CHIME/Pulsar instrument hardware is funded by the Natural Sciences and Engineering Research Council (NSERC) Research Tools and Instruments (RTI-1) grant EQPEQ 458893-2014.
The work of DAZ and AVK was supported by the Russian Science Foundation, 
grant number 22-22-00921, \url{https://rscf.ru/project/22-22-00921/}.
DAZ thanks Pirinem School of Theoretical Physics for hospitality. 
SVZ acknowledges PAPIIT grant IN119323.
Pulsar research at UBC is funded by an NSERC Discovery Grant and by the Canadian Institute for Advanced Research. F.~A.~D. is supported by the UBC Four Year Fellowship. M.~A.~M. is supported by the NANOGrav NSF Physics Frontiers Center award numbers 1430284 and 2020265 and NSF award number 2009425. J.~K.~S. is supported by the NANOGrav NSF Physics Frontiers Center award numbers 1430284 and 2020265.
This work has made use of data from the European Space Agency (ESA) mission
{\it Gaia} (\url{https://www.cosmos.esa.int/gaia}), processed by the {\it Gaia}
Data Processing and Analysis Consortium (DPAC,
\url{https://www.cosmos.esa.int/web/gaia/dpac/consortium}).
Funding for the DPAC
has been provided by national institutions, in particular the institutions
participating in the {\it Gaia} Multilateral Agreement.
The Pan-STARRS1 Surveys (PS1) and the PS1 public science archive have been made possible through contributions by the Institute for Astronomy, the University of Hawaii, the Pan-STARRS Project Office, the Max-Planck Society and its participating institutes, the Max Planck Institute for Astronomy, Heidelberg and the Max Planck Institute for Extraterrestrial Physics, Garching, The Johns Hopkins University, Durham University, the University of Edinburgh, the Queen's University Belfast, the Harvard-Smithsonian Center for Astrophysics, the Las Cumbres Observatory Global Telescope Network Incorporated, the National Central University of Taiwan, the Space Telescope Science Institute, the National Aeronautics and Space Administration under Grant No. NNX08AR22G issued through the Planetary Science Division of the NASA Science Mission Directorate, the National Science Foundation Grant No. AST-1238877, the University of Maryland, Eotvos Lorand University (ELTE), the Los Alamos National Laboratory, and the Gordon and Betty Moore Foundation.

\section*{Data Availability}

The optical data are available through the GTC data 
archive: \url{https://gtc.sdc.cab.inta-csic.es/gtc/}, the CHIME data and 
the \eros\ data -- upon request.\\~\\

\bibliographystyle{mnras}
\bibliography{ref}



\bsp	
\label{lastpage}
\end{document}